\documentclass{article}[12pt]
\usepackage{graphicx,tikz,endnotes,amsmath,bbm}
\usepackage[natbibapa]{apacite}
\usetikzlibrary{bayesnet}
\usepackage{multirow}
\usepackage[flushleft]{threeparttable}
\usepackage{float}
\usepackage[labelfont=bf]{caption}
\usepackage{geometry}
\usepackage[doublespacing]{setspace}
\usepackage[toc]{appendix}
\usepackage{longtable,siunitx}

\let\footnote=\endnote

\usepackage{titlesec}

\titleformat{\section}
{\large\bfseries}{}{0em}{}
\titleformat{\subsection}
{\large\bfseries\itshape}{}{0em}{}

\usepackage{hyperref}
\hypersetup{
	linkcolor  = blue,
	citecolor  = blue,
	urlcolor   = blue,
	colorlinks = true,
}

\title{\Large{Election campaigning on social media: Politicians, audiences and the mediation of political communication on Facebook and Twitter}}

\date{}
\author{Sebastian Stier, Arnim Bleier, Haiko Lietz, Markus Strohmaier \\\emph{GESIS -- Leibniz Institute for the Social Sciences}\\\href{mailto:sebastian.stier@gesis.org}{\normalsize \texttt{sebastian.stier@gesis.org}}}

\begin{document}

\newcommand{\superscript}[1]{\ensuremath{^{\textrm{#1}}}}

\setlength{\abovedisplayskip}{2pt}
\setlength{\belowdisplayskip}{2pt}


\maketitle
\begin{abstract}
Although considerable research has concentrated on online campaigning, it is still unclear how politicians use different social media platforms in political communication. Focusing on the German federal election campaign 2013, this article investigates whether election candidates address the topics most important to the mass audience and to which extent their communication is shaped by the characteristics of Facebook and Twitter. Based on open-ended responses from a representative survey conducted during the election campaign, we train a human-interpretable Bayesian language model to identify political topics. Applying the model to social media messages of candidates and their direct audiences, we find that both prioritize different topics than the mass audience. The analysis also shows that politicians use Facebook and Twitter for different purposes. We relate the various findings to the mediation of political communication on social media induced by the particular characteristics of audiences and sociotechnical environments.
\\[0.2cm]
Forthcoming in \emph{Political Communication}\\
DOI: \href{https://doi.org/10.1080/10584609.2017.1334728}{https://doi.org/10.1080/10584609.2017.1334728}\\[0.5cm] 
\end{abstract}


\section{Introduction}
Social media have become ubiquitous communication channels for candidates during election campaigns. Platforms like Facebook and Twitter enable candidates to directly reach out to voters, mobilize supporters and influence the public agenda. These fundamental changes in political communication therefore present election candidates with a widened range of strategic choices. Should candidates address the topics most important to a mass audience? Should they tailor their messages to the specific habits and audiences on social media platforms? Although academic research on social media campaigning has flourished in the last several years \citep{Jungherr.2016a,Boulianne.2016}, it is still unclear which topics politicians address on these platforms, since previous research mostly concentrated on meta data generated by the use of communication conventions such as retweets, @-mentions, likes or hashtags. Understanding the ways in which politicians adapt the contents of their messages to the peculiarities of different platforms generates deeper insights into how political communication is shaped by social media.

Much research revealed a continuation of the status quo in online campaigning, as politicians mostly replicated traditional messages and campaign modes on their web presences while limiting engagement with users \citep{Stromer.2000,Lilleker.2011,Gibson.2014,Larsson.2015}. New media notwithstanding, it seems imperative for candidates to tailor their online messages towards the preferences of the majority of voters, in line with models of mass communication \citep{Downs.1957,Druckman.2010}. However, the observation of \citet[p.~140]{Mcquail.2010} that ``the audience member is no longer really part of a mass, but is either a member of a self-chosen network or special public or an individual'' is especially true for interactive social media. On these platforms, politicians get directly exposed to users with rather specific demographic characteristics and political interests \citep{Diaz.2016,Nielsen.2013, Schoen.2013} and have to adapt to unique affordances of social media sites \citep{ Hoffmann.2017,Jungherr.2016bb}. Candidates might therefore tailor their communication to the sociotechnical environments of platforms like Facebook and Twitter. To infer if campaign messages are aimed at a mass audience or more particular sets of audiences, we analyze ``the distribution of salience across a set of issues'' \citep[p.~396]{Iyengar.1979} and make a novel contribution by (1) deriving these distributions from survey responses to proxy the preferences of a mass audience, as well as (2) from different social media and (3) from multiple content layers (politicians and audiences on social media).

The empirical analysis focuses on political communication on Facebook and Twitter by candidates during the German federal election (\emph{Bundestagswahl}) campaign 2013. The baseline is a representative survey of the German population conducted during the election campaign in which participants were asked to describe in free text the most important contemporary political problem. These responses were assigned to 18 topic categories by the survey administrators. We present a human interpretable Bayesian language model which allocates social media messages to the known topic categories from the survey responses, but creates additional social media specific topics, if necessary. Applying the model to social media messages of candidates and their social media audiences, we find that although their topic agendas converge to some extent with the survey, both prioritize topics like campaign-related events that are different from the concerns of a mass audience. Furthermore, the focus of candidates on social media is more similar to the topics discussed by the audiences to which they are most directly exposed. An analysis of the language used in messages adds methodological robustness to the results and confirms that politicians primarily use Facebook for campaign-related purposes like the promotion of their activities while preferring Twitter to comment on contemporary political events. We relate these differences to the mediation of political communication on social media induced by diverging characteristics of audiences and sociotechnical environments.

\section{Related literature and research gaps}
\label{sec:theory}
Our study is located at the intersection of cross-media and social media research. There is an established research tradition relating the use of different media to outcomes and processes like political knowledge, participation and voting \citep{Prior.2007}, news consumption \citep{Althaus.2002} and political communication \citep{Druckman.2010}. 

In terms of election campaigning on the Internet, \citet{Druckman.2010} presented several relevant findings for our study. The campaign officials the authors surveyed revealed that even though they were aware that supporters are the most frequent visitors of candidate websites, these formats were still designed for a mass audience. In a comparison of websites and TV ads, the authors showed that candidates are equally likely to use both media for negative campaigning, implying that the medium and different user groups do not matter much in campaign strategy. Other studies mirrored the rather conservative use of the web by politicians (e.g. \citealp{Stromer.2000,Lilleker.2011,Gibson.2014,Larsson.2015}). 

In terms of audience behavior, \citet{Althaus.2002} established that whether news are consumed offline or online affects perceptions of issue importance. The selection of news at the individual level is ``filtered'' by different media, even when the original source is the same. While the newspaper edition of the New York Times heavily guides the readers by providing journalistic cues and a steady diet of ``public affairs coverage'', users of its online edition were navigating contents in a manner that suits their personal preferences. Others have observed an audience fragmentation and turn towards entertainment formats in high-choice media environments as well (e.g. \citealp{Prior.2007,Nielsen.2013}). 

Taken together, these findings posit that although the citizens who actually use the Internet for political purposes have rather specific political interests, politicians have so far used the web with a mass audience in mind. The skewed perceptions of topic importance by Internet users that \citet{Althaus.2002} observed should be even more pronounced when using social media, given the social and algorithmic cues that these platforms provide. Accordingly, \citet{Jungherr.2016bb} showed that during the German federal election campaign 2013, topic priorities of Twitter audiences deviated from a survey and mass media coverage. Strategic campaigns should adapt to these environments and narrowcast their messages to the particular audiences they encounter \citep{Kreiss.2016}.

Election campaigning on social media has been studied extensively, as researchers examined how election campaigns unfold, how candidates are embedded in communication networks and how they interact among themselves and with the public (cf. \citealp{Jungherr.2016a,Boulianne.2016}). Still, in terms of cross-media research, this literature is limited in several regards: first, most studies focused on one isolated platform, overwhelmingly Twitter and -- less often -- Facebook. Second, only a fraction of this work concentrated on the actual contents of communication going beyond meta data, i.e., digital traces left by communication artifacts like @-messages, retweets, likes or hashtags. While several studies coded contents of social media posts by U.S. parliamentarians \citep{Golbeck.2010,Gainous.2014,Bronstein.2013}, these efforts mostly consisted of smaller samples and/or did not specifically categorize the topics politicians talk about. Third, most research is confined to the boundaries of election campaigning on a given social media platform. The few cross-platform analyses either restricted themselves to main accounts of party organizations \citep{Rossi.2016,Larsson.2015} or metrics of attention like the number of views or followers on multiple platforms \citep{Nielsen.2013}. In recent advances, \citet{Karlsen.2016} linked a candidate survey with Twitter data to uncover candidates' strategies and determinants of Twitter success. \citet{Bode.2016} compared television advertisements with Twitter data and identified deviations indicating that the two media represent distinct modes of campaigning. Building on that, a comparison between multiple social media platforms might reveal even more fine-grained affordances of different media. Such platform specific mediation effects urgently need to be taken into account in models of political communication \citep{Jungherr.2016bb}. 

We add to this research by integrating features from multiple spheres of political communication in one research design. Recent advances in quantitative text analysis \citep{Grimmer.2013} enable us to perform text analysis at a larger scale. We regard responses to an open-ended question in a representative survey as reflections of the topic priorities of a mass audience and take these as an empirical base for the analysis of social media messages by candidates and their audiences.

\section{Strategic election campaigning on social media}

Politicians seeking election need to be responsive to the political preferences of their constituencies \citep{Downs.1957}. However, it is an open question if politicians tailor their online messages to the topic priorities of a mass audience or particular social media audiences. In contrast to \citet{Druckman.2010} who revealed rather traditional strategies on campaign websites, we argue that social media poses a yet again different communication constellation: politicians are embedded in an interactive context which skews their messages to the topic preferences of their immediate communication network (see also \citealp{Bode.2016}). This might be due to strategic reasoning in order to increase the success of messages or an unwitting outcome of the uses and gratifications of politicians themselves. 

Our empirical study relies on survey and social media generated during the German federal election campaign 2013. Given the German electoral system where party identification is still rather strong \citep{Arzheimer.2006}, district-level topics are of minor importance in election campaigns and public agendas mostly converge to the topics salient at the national level. Therefore, topic salience expressed in politicians' social media messages can reasonably be compared to topic salience in public opinion polls. To the best of our knowledge, no such research has been undertaken so far, which naturally makes our study an exploratory one. Yet we derive useful propositions from previous research to develop several theoretical expectations.

\subsection{Social media as part of multifunctional online campaigns}

When theorizing on the topics of politicians' campaign messages, at least three arguments indicate that social media indeed pose campaign environments distinct from mass communication arenas. First, we have to consider that social media are not only used to address political topics important to a mass audience, but perform several other functions in election campaigns. \citet{Kobayashi.2015}, for instance, identified three functions: promoting issue positions, demonstrating beneficial personality traits and improving name recognition. \citet{Jungherr.2016zz} proposed a four-fold typology distinguishing between organizational uses, active campaigning in information spaces, resource collection and allocation as well as symbolic purposes. A considerable share of online campaigning should therefore be devoted to the mobilization of supporters, organization of campaigns \citep{Lilleker.2011,Nielsen.2013} and representational/symbolic purposes. Contrary to the ``electronic brochures'' of websites \citep{Stromer.2000,Hoffmann.2017}, a campaign has to internalize a whole set of platform-specific affordances on social media in order to demonstrate that it represents the ``state of the art''.

Second, the demographic composition as well as the political preferences and interests of social media audiences are much different from a representative sample of citizens \citep{Diaz.2016,Schoen.2013}. In the context of the German election 2013, political audiences on Twitter rarely talked about core political issues like the Euro crisis or energy policy, but predominantly addressed NSA surveillance and campaign-related events like the televised debate between the leading national candidates \citep{Jungherr.2016bb}. Strategic, ``micro-targeted'' \citep{Kreiss.2016} campaigns should tailor their messages to specific audiences and successful marketing strategies on social media. While the German political system and privacy regulations certainly pose barriers to data-driven micro-targeting \citep{Stier.2015c}, mediation effects of social media platforms should still be felt in online campaigning. In consequence, candidates might be inclined to address topics like internet policy, which are more important to social media than mass audiences, and unfolding campaign events. 

Third, candidates themselves have different uses and gratifications of online media \citep{Hoffmann.2017,Marcinkowski.2014}. One of the central features of the Internet is its coalescence of private and public functions \citep[p.~41]{Mcquail.2010}, and many German politicians indeed operate their Twitter accounts personally \citep{spiegel.2015}. Studying individual-level predictors of online campaigning, \citet{Marcinkowski.2014} revealed that German state-level candidates with a more positive attitude towards the Internet use its applications more intensively. \citet{Hoffmann.2017} reported that although the motives self-promotion and information dissemination are dominant among Swiss politicians, which is consistent with traditional modes of campaigning, they also use social media for more personal uses like information seeking and entertainment. The authors conclude that ``specific motives---possibly based on varying levels of understanding of and experience with new media---can lead to more or less avid and strategic ICT adoption'' \citep[p.~251]{Hoffmann.2017}. 

Given the multiple functions of campaigns as well as an adoption of typical practices and engagement with specific audiences by a considerable share of candidates, it can be assumed that the topics salient in politicians' social media messages do not necessarily reflect the topic priorities of a mass audience. 

\textbf{H1:} Topic saliences in messages by candidates and audiences on social media are more similar to each other than to topic salience among a mass audience.

\subsection{The use of Facebook and Twitter by election candidates}

Due to the different architectures of social media platforms, topic saliences could also differ between the two networks we look at, Facebook and Twitter. Previous studies have shown that different media logics influence the strategic considerations underlying election campaigns \citep{Kreiss.2016,Bode.2016}. On Twitter, most user accounts are publicly visible and accessible even for non-registered audiences. Its usage is centered around topics and the retweet feature facilitates the diffusion of political information beyond the direct follower network via two-step flow processes \citep{Wu.2011}. In contrast, most accounts on Facebook are private and its usage is based on one-way or reciprocal friendship ties. Information travels less fluidly through this medium, also due to the extensive algorithmic filtering of contents. Thus, the audience for Facebook posts mostly consists of people already ``liking'' a candidate page (see for similar audience-centered arguments: \citealp{Norris.2003,Nielsen.2013}). 

Among the most active Twitter users are prime targets of campaigns like political elites and influentials. Journalists, for instance, regard Twitter of higher value for news reporting while using Facebook primarily for private purposes \citep{Parmelee.2014}. Campaign officials interviewed by \citet{Kreiss.2016} emphasized that journalists use Twitter as an ``index of public opinion'', which implies that targeted campaign messages on Twitter had the potential to create spillover effects to other media. Similarly, Twitter is used more intensively by Swedish political parties than Facebook due to the greater potential to reach opinion leaders via the former medium \citep{Larsson.2015}. Disaggregating campaigning by political arena, \citet{Larsson.2016b} provided evidence from Norway that local politicians valued Facebook higher for political communication, whereas politicians at the national level preferred Twitter. The former medium seems to be particularly suited for local politics, which election campaigning is to a large extent about, while the latter is being used to connect to national audiences.

Considering these findings, we expect that candidates use Twitter to contribute to contemporary national debates about policies or high-attention campaign events like TV debates. In contrast, politicians ``preach to the converted'' on Facebook (using the words of \citealp{Norris.2003}), mostly party supporters and local constituents. Since these users already show a considerable interest in a candidate and her political stances, the strategic value of this friendly audience lies in mobilization, i.e., persuading followers to take part in the election campaign as volunteers or turn out to vote, rather than in convincing them of policy propositions.

\textbf{H2:} Topic salience among a mass audience is more similar to topic salience by candidates on Twitter than to topic salience by candidates on Facebook.

\textbf{H3a:} Candidates prefer Twitter to discuss policies and unfolding campaign events like TV debates. 

\textbf{H3b:} Candidates prefer Facebook for campaigning and mobilization purposes.

\section{Methodology}\label{sec:meth}
\textbf{Language model.}
We begin with some considerations regarding the joint modeling of survey responses and social media messages. Classification is a supervised task of identifying to which group, here topics, a document (a survey response, Facebook message or tweet) belongs on the basis of labeled training data. Clustering is an unsupervised task of grouping observations that are more likely to follow the same word distributions as those in other groups \citep{bishop2006pattern}. Our model occupies a middle ground between classification and clustering. We use labeled survey responses as training data, but then cluster those social media messages that use distinct words or word combinations than survey responses into new topics that capture the specifics of online communication.

To this end, we introduce a Bayesian semi-supervised nonparametric single-membership language model. The model allows for the grouping of short messages into known survey classes and, if necessary, additional social media specific topics. The underlying assumptions of the model are that documents are generated in a probabilistic process of (1) drawing a topic for a document and then (2) drawing the words of that document from the selected topic.
Such a content analysis identifies differences in language as practiced culture and is therefore particularly suited to distinguish the different communication styles of survey and social media contexts \citep{McFarland.2013}. 

The model is semi-supervised, allowing for the incorporation of survey responses coded with a topic label. The labeled documents are responses to an open-ended survey question on the most important contemporary political problem in Germany (MIP). Moreover, the model is nonparametric, i.e., we do not set the total number of topics in advance \citep{teh2010}. Instead, we use a Dirichlet process prior on the number of topics. While the allocation of survey responses to topics is known, the allocation of social media messages to topics is unobserved and has to be inferred. The nonparametric prior allows for new topics specific to Facebook and Twitter communication to be established when messages cannot plausibly be explained by the topics used in the survey. Finally, we follow the single-membership assumption that each document is generated from a single topic. This assumption handles short documents such as social media messages or survey responses well \citep{yin2014dirichlet}. In fact, the short length of documents effectively prohibits a mixed-membership approach, such as the one put forward by \citet{Roberts.2013x}. A formal description of this model, specifically tailored to our needs, is given in Appendix \ref{appendix:model}.

The inference step of the model described in Appendix \ref{appendix:model_inference} reverses assumptions for document generation to an automated reading of otherwise latent semantic topics from the social media datasets. For this, we ran the Gibbs sampling procedure for 100 iterations. For the evaluation of model outputs, the most important criterion was the substantive fit as determined by human judgment, i.e., the quality of topics in the context of our research question \citep[p.~286]{Grimmer.2013}. As parameters, we chose $\alpha = 1.0$ and $\beta = 1.5$ which produce well-interpretable topics that cluster messages unique to social media while allocating topically related social media messages to the known survey topics. The final output provides an aggregated and unified macroscopic view on signals in political communication that would be impossible to gather with conventional methods.

\textbf{GLES survey data.}
The MIP we use was administered in the Rolling Cross-Section of the German Longitudinal Election Study (GLES) \citep{gles.2013} accompanying the federal election 2013.\footnote{The wording is: ``In your opinion what is the most important political problem facing Germany at the moment?'' \citep{gles.2013}.} The use of such an MIP has been criticized in terms of conceptualization and measurement in previous research. Some respondents relate their answer to their individual situation, some to the political situation at the societal level, and some consider the question based on their voting considerations \citep{Wlezien.2005}. People therefore have different understandings of topic importance, of the extent to which a topic should be considered as a problem and also of the perspective that should be taken when making these judgments. These methodological considerations notwithstanding, our application of the MIP is less problematic than its predominant application in survey research, in which open-ended responses are reduced to one quantitative scale of issue salience. In contrast, we take the entire textual contents of each response into account and therefore capture all the semantic facets related to the \emph{importance} of topics and citizens' perception of them as being \emph{problematic} (arguing similarly: \citealp{Mellon.2014}). Therefore, the MIP question serves our methodology well as a ``catch all category'' soaking up the nuances in the vocabulary used by the respondents, which will in turn also be reflected in the inference stage when the model is applied to social media messages.
 
The GLES survey was in the field from 8 July until 3 November 2013, surveying 7,882 unique participants twice, in a pre-election and a post-election wave. We pooled all responses on the first and second MIP during this time period, which gives us a total of 23,604 observations.\footnote{Before the election, 7,249 participants (92\%) stated a MIP and 6,673 (85\%) stated a second MIP. After the election, 5,016 respondents (64\%) stated a MIP and 4,666 (59\%) stated a second MIP.}
Pooling implicates that observations are not independent of each other, since most participants responded multiple times in the survey.\footnote{4,220 (54\%) stated four problems, 543 (7\%) stated three, 2,364 (30\%) stated two and 367 (5\%) stated one problem. 388 (5\%) did not know a problem or did not respond.} Thus, the sample loses its initial representativeness when we aggregate responses to GLES topic categories for our empirical analysis. Yet the practical benefits of pooling to construct a sufficiently large training dataset outweigh its theoretical disadvantages, since we primarily use the survey responses for the extraction of textual information.

The open-ended responses were coded by GLES experts according to a hierarchical categorization of topics \citep{GLES.2013.agendafragen} which consists of the three traditional higher-level dimensions used in political science (politics, polity and policy). 188 responses (0.8\%) were concerned with political processes (politics) while 1,063 observations (4.5\%) were concerned with political structures (polity). The overwhelming share of survey responses thus located the most important problem in a policy area. The classes for these responses required additional manual filtering and reassembling in order to arrive at discriminative and more equally sized training classes. As a result, we obtained 18 topics for training (see Appendix \ref{appendix:survey}).

\textbf{Social media data.}
We use a dataset covering political communication on Facebook and Twitter during the election campaign for the German Bundestag 2013 gathered via the Facebook Graph API and the Twitter Streaming API \citep{Kaczmirek.2013}. From this dataset, we extracted all social media messages posted by candidates, their incoming @-mentions on Twitter and the comments to their posts on Facebook during the research period analogous to the survey.\footnote{The data mining of Facebook posts ended three days earlier, on 31 October 2013.} We only selected posts by parties with a realistic chance of passing the electoral threshold of 5\% required for representation in the Bundestag, namely CDU, CSU, FDP, Gr\"{u}ne, Linkspartei and SPD. The AfD almost made it into the Bundestag in 2013, however, its rise to prominence in public opinion polls came late and thus the party was not incorporated in the data collection.

The dataset contains 49,573 Facebook posts and 134,462 tweets by candidates. As audience data, we preselected all 180,214 comments on candidates' Facebook posts and all 282,118 tweets that @-mentioned at least one candidate. These numbers at first seem to reflect a low interest by the public. However, it should be kept in mind that election-related topics are also discussed without commenting on candidates' Facebook posts or mentioning them in tweets. We chose these two particular metrics since they are the most direct exposure of a candidate to messages by her audience (account holders per default get a notification via the platform interface and via email that they received a mention/comment).

\textbf{Preprocessing.}
We preprocessed our five corpora (survey, politicians Facebook, politicians Twitter, audience Facebook, audience Twitter) before we applied our model, which is a necessary reduction of linguistic complexity for text analysis \citep{Grimmer.2013}. The preprocessing of documents involved removing punctuation, standard German stop words, URLs and Twitter user handles.
In addition, we removed the names of all political parties and candidates to prevent inferring topics that are party-specific or centered around social relations.
Customized stop words, compiled inductively while refining the model, were also removed. This customized stop word list contains ambivalent words with high probabilities in many distinct topics (e.g., ``Politik'') as well as very frequent and very infrequent words. Finally, only those social media messages were kept that contain at least three words in order to incorporate sufficient information for the clustering procedure of the model. Survey responses were allowed to consist of only one word since the topic was already known for these documents. 

In addition to text cleaning, we made sure that the four social media corpora are comparable in size and number of words used so that they have a comparable influence on topic construction. To achieve this, we operationalized the Facebook and Twitter audience corpora as random samples of equal size to politicians' messages, stratified by political party. For instance, the 5,902 tweets by FDP candidates were mirrored by the same number of audience tweets mentioning an FDP candidate and their 1,777 Facebook posts were matched by as many randomly sampled accompanying Facebook comments.\footnote{We took audience tweets in which only one politician was mentioned for the sampling.}
In total, 22,186 survey responses, 17,546 Facebook posts by politicians, 17,546 Facebook comments, as well as 54,093 tweets by politicians and 54,093 @-mentions were used as inputs for the model.

\section{Results and discussion}

\subsection{Analysis at the document level}

\begin{figure}
	\centering	
	\includegraphics[width=0.8\textwidth]{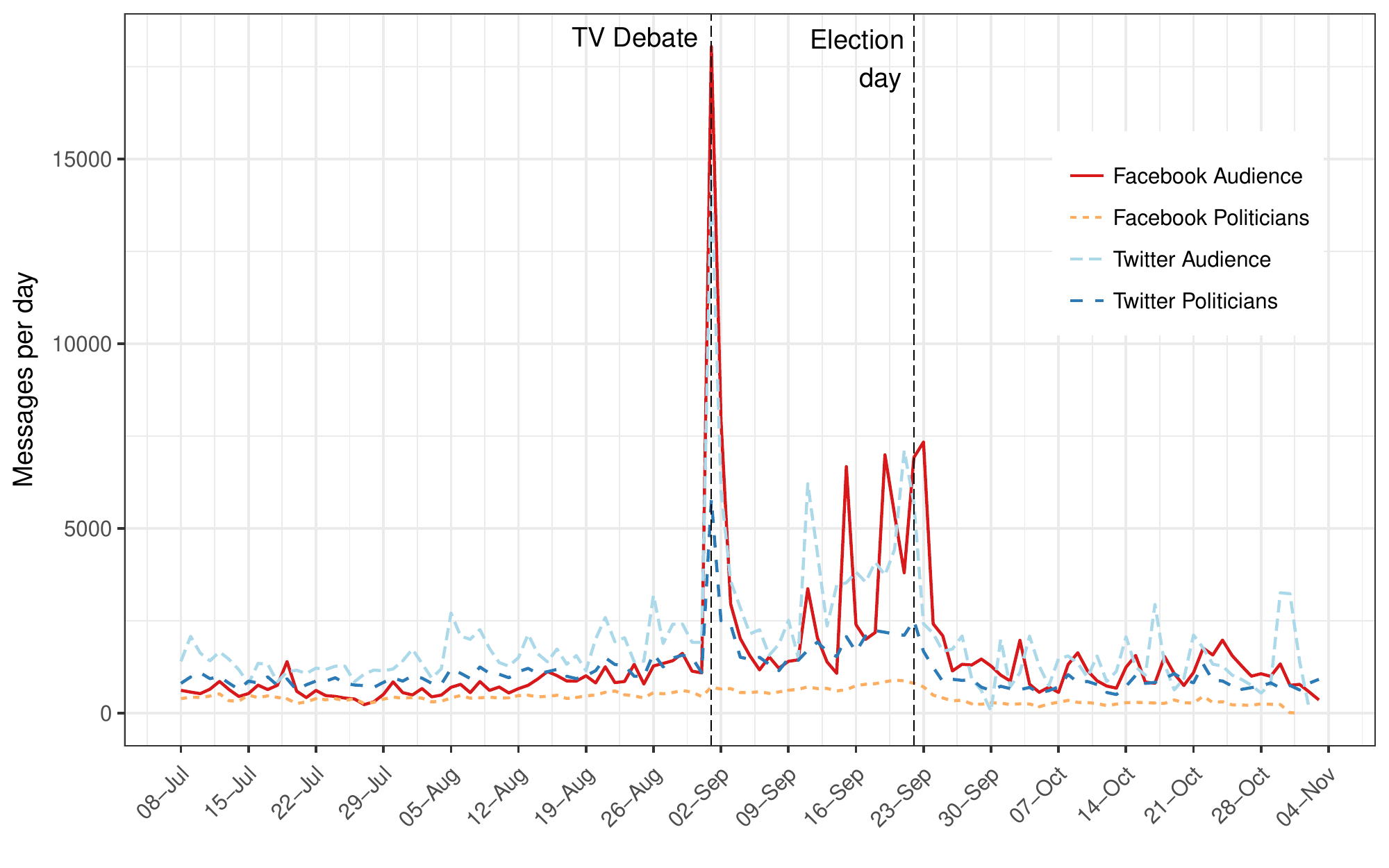}
	\caption{Social media messages over time.}
	\label{fig:socialmedia_timeseries}
\end{figure} 

We start our analysis with a description of social media activity by politicians and audiences over time using the raw counts of messages from the unpreprocessed dataset. This gives us first indications about specific focus points and the topics predominantly addressed on social media. The vertical lines in Figure \ref{fig:socialmedia_timeseries} represent the TV debate between the party leaders Angela Merkel and Peer Steinbr\"{u}ck (1 September 2013) and election day (22 September 2013). Especially the televised debate drew attention (see also \citealp{Jungherr.2016bb,Lietz.2014}), since social media users are particularly active during these high-attention periods \citep{Diaz.2016,Kreiss.2016}. Interestingly, the Facebook politicians time series barely reacts to the TV debate and not nearly in the order of magnitude like the other three time series. This indicates that politicians prefer Twitter over Facebook to comment on unfolding events (supporting H3a). We will follow up on these hints with an extensive analysis of the topics discussed by politicians and their audiences on social media.

The text analysis model described above generates 51 topics; the 18 known topics from the survey and 33 additional ones based on the social media corpora. We removed 23 new topics with less documents than the smallest known topic from the survey ($<468$ documents, aggregated across all social media corpora) in order to improve interpretability. Those 23 residual topics in total only covered 1,387 of 143,278 social media documents. The first 18 topics were labeled using the known topic labels from GLES (cf. Table \ref{tab:topics_gles}). The ten remaining social media topics were labeled by inspecting the documents for each topic qualitatively. The few ambiguous cases were discussed among the authors and decided upon consensually.

\renewcommand{\arraystretch}{0.6}
 
\begin{longtable}{p{.975\textwidth}}
    \caption{Top words per topic}
    \label{tab:topics} \\ \hline

\emph{Known topics from the survey} \\ \hline
\footnotesize \textbf{Budget \& Debt}: schulden, schuldenabbau, staatsverschuldung, verschuldung, haushalt, finanzen, ueberschuldung, staatsschulden, abbau, neuverschuldung \emph{(debt, debt reduction, national debt, debt, budget, finances, debt overload, national debt, reduction, new debt)} \\[0.2cm]
\footnotesize \textbf{Currency \& Euro}: euro, eurokrise, schulden, europa, griechenland, krise, geld, milliarden, banken, schuldenkrise \emph{(euro, euro crisis, debt, europe, greece, crisis, money, billions, banks, debt crisis)} \\[0.2cm]
\footnotesize \textbf{Economy}: wirtschaft, finanzkrise, wirtschaftliche, wirtschaftspolitik, wirtschaftskrise, stabilitaet, wirtschaftlichen, banken, versteht, wachstum \emph{(economy, financial crisis, economic, economic policy, economic crisis, stability, economic, banks, understand, growth)} \\[0.2cm]
\footnotesize \textbf{Education}: bildung, bildungspolitik, mitte, schulen, jaehrige, schule, geld, lehrer, kinder, schueler \emph{(education, education policy, middle, schools, annual, school, money, teacher, children, pupils)} \\[0.2cm]
\footnotesize \textbf{Environment}: umwelt, umweltschutz, umweltpolitik, wohlstand, klimawandel, klima, klimaschutz, aufbruch, oekologie, energiewende \emph{(environment, environmental protection, environmental policy, prosperity, climate change, climate, climate protection, start, ecology, energy transformation)} \\[0.2cm]
\footnotesize \textbf{Family Policy}: kinder, kita, frauen, betreuungsgeld, familie, familienpolitik, eltern, familien, gleichstellung, kind \emph{(children, kita, women, child care subsidy, family, family policy, parents, families, equalization, child)} \\[0.2cm]
\footnotesize \textbf{Foreign Policy (Defense)}: syrien, fluechtlinge, krieg, lampedusa, frieden, russland, aussenpolitik, muessen, europa, waffen \emph{(syria, refugees, war, lampedusa, peace, russia, foreign policy, must, europe, weapons)} \\[0.2cm]
\footnotesize \textbf{Foreign Policy (Europe)}: europa, europapolitik, europaeische, europaeischen, zusammenhalt, integration, europas, stabilitaet, laender, euro \emph{(europe, european policy, european, solidarity, integration, europe's, stability, countries, euro)} \\[0.2cm]
\footnotesize \textbf{General Fiscal Policy}: finanzen, finanzpolitik, finanzielle, finanzlage, sicherheit, geld, finanzmarkt, finanzierung, situation, ordnung \emph{(finances, fiscal policy, financial, financial situation, security, money, capital market, funding, situation, order)} \\[0.2cm]
\footnotesize \textbf{General Social Policy}: soziale, gerechtigkeit, altersarmut, reich, armut, schere, ungerechtigkeit, sozialen, wandel, sozialpolitik \emph{(social, justice, poverty of the elderly, rich, poverty, gap, injustice, social, change, social policy)} \\[0.2cm]
\footnotesize \textbf{Health Care \& Pensions}: renten, rente, pflege, buergerversicherung, gesundheitspolitik, tvduell, rentenpolitik, gesundheit, gesundheitswesen, medizin \emph{(pensions, pension, caregiving, citizen insurance, health policy, tvduell, pension policy, health, health sector, medicine)} \\[0.2cm]
\footnotesize \textbf{Infrastructure}: energiewende, energie, erneuerbare, strom, energien, kohle, energiepolitik, volksentscheid, umlage, klimaschutz \emph{(energy transformation, energy, renewable, electricity, energies, coal, energy policy, referendum, contribution, climate protection)} \\[0.2cm]
\footnotesize \textbf{Labor Market}: mindestlohn, arbeitslosigkeit, arbeit, euro, muessen, steuern, geld, leben, maut, tvduell \emph{(minimum wage, unemployment, work, euro, must, taxes, money, live, toll, tvduell)} \\[0.2cm]
\footnotesize \textbf{Law \& Order}: nsa, prism, snowden, ueberwachung, datenschutz, affaere, bnd, freiheit, daten, buerger \emph{(nsa, prism, snowden, surveillance, data protection, affair, bnd, freedom, data, citizens)} \\[0.2cm]
\footnotesize \textbf{Migration \& Integration}: integration, auslaender, auslaenderpolitik, zuwanderung, migranten, einwanderung, asylbewerber, asylanten, migration, einwanderungspolitik \emph{(integration, foreigners, policy towards foreigners, immigration, migrants, asylum seeker, migration, asylum policy)} \\[0.2cm]
\footnotesize \textbf{Politics}: euro, muessen, buerger, jahr, unternehmen, arbeit, stadt, land, region, leben \emph{(euro, must, citizens, year, businesses, work, city, country, region, live)} \\[0.2cm]
\footnotesize \textbf{Polity I}: geld, ehrlichkeit, glaubwuerdigkeit, buerger, bevoelkerung, demokratie, uneinigkeit, ausland, volk, vertrauen \emph{(money, honesty, credibility, citizens, population, democracy, disagreement, foreign countries, people, trust)} \\[0.2cm]
\footnotesize \textbf{Taxes}: steuern, steuerpolitik, progression, steuererhoehung, steuer, kalte, kalten, abbau, steuergerechtigkeit, steuererhoehungen \emph{(taxes, fiscal policy, progression, tax increase, tax, cold, reduction, fiscal justice, tax increases)} \\ \hline
\emph{New topics found on social media} \\[0.2cm] \hline
\footnotesize \textbf{Campaigning (Events)}: veranstaltung, talk, podiumsdiskussion, wahlkreis, gast, einladung, interview, unterwegs, bundestagswahl, gespraeche \emph{(event, conversation, panel discussion, constituency, guest, invitation, interview, on the road, federal election, talks)} \\[0.2cm]
\footnotesize \textbf{Campaigning (Local)}: danke, dank, super, guten, stimmung, spass, infostand, wahlkreis, unterwegs, aktion \emph{(thanks, thank, super, good, atmosphere, fun, info booth, constituency, on the road, action)} \\[0.2cm]
\footnotesize \textbf{Coalition Formation}: koalition, grosse, mehrheit, waehler, waehlen, grossen, muessen, bundestag, demokratie, opposition \emph{(coalition, grand, majority, voter, vote, grand, must, bundestag, democracy, opposition)} \\[0.2cm]
\footnotesize \textbf{Demonstrations}: nazis, demo, baden, wuerttemberg, angst, innen, wasser, freiheit, platz, protest \emph{(nazis, demo, baden, wuerttemberg, fear, inside, water, freedom, square, protest)} \\[0.2cm]
\footnotesize \textbf{Misconduct}: paedophilie, zurueck, treten, debatte, verfassungsschutz, tritt, unfassbar, arbeit, ruecktritt, aufloesen \emph{(pedophilia, back, step, debate, internal intelligence service, step, inconceivable, work, resignation, close)} \\[0.2cm]
\footnotesize \textbf{NSA Surveillance}: snowden, edward, moskau, treffen, trifft, brief, nsa, respekt, germany, russland \emph{(snowden, edward, moscow, meeting, meet, letter, nsa, respect, germany, russia)} \\[0.2cm]
\footnotesize \textbf{Parliamentary Procedures}: bundestag, nsu, sitzung, fraktion, bundestages, fraktionssitzung, landesgruppe, bundestagsfraktion, rheinland, rede \emph{(bundestag, nsu, session, faction, bundestag, caucus, regional group, bundestag faction, rhineland, speech)} \\[0.2cm]
\footnotesize \textbf{Political Debates}: tvduell, waehlen, danke, bayern, dreikampf, zeit, richtig, beide, duell, kanzlerin \emph{(tvduell, vote, thank, bavaria, triathlon, time, right, both, duel, chancellor)} \\[0.2cm]
\footnotesize \textbf{Polity II}: leben, muessen, land, geld, buerger, wissen, richtig, volk, kinder, freiheit \emph{(live, must, country, money, citizens, know, right, people, children, freedom)} \\[0.2cm]
\footnotesize \textbf{Post Election}: glueckwunsch, herzlichen, bundestag, ergebnis, erfolg, geburtstag, dank, danke, nsa, gewaehlt \emph{(congratulation, cordial, bundestag, result, success, birthday, thank, thanks, nsa, elected)} \\ \hline

\end{longtable}

Table \ref{tab:topics} lists the ten most probable words per topic in German and English. The topics are very well interpretable. Overall, the model successfully assigned the social media documents to the known classes when appropriate. With the exception of \emph{NSA Surveillance} (which is assigned to \emph{Law \& Order} in the survey), no new topic mirroring an existing one was created. An inspection of the documents shows that the new topic focuses on Edward Snowden and Russia, while the topic from the survey concentrates on the implications of NSA spying for citizens. One new topic, \emph{Polity II}, revolves around the role of citizens in, and criticism of, Germany's political structure. The topic is distinct from \emph{Polity I} in the survey, which is more about the role of politicians in the polity. Besides these two topics, all other new topics are related to political processes. Campaign events leave particularly strong marks in the topic \emph{Political Debates}, which is the largest new topic and also the most ambiguous one, as it contains a great variety of exchanges between politicians and audiences. Especially the televised debate is prominently featured, which concurs with \citet{Kreiss.2016} who showed that Twitter is used by campaigns for the engagement with audiences and journalists during high-attention events.

\begin{table}[t]
    \centering
    \caption{Topic salience per corpus (in percent)}
    \label{tab:fraction}
   \begin{tabular}{lccccc}

 &  & \multicolumn{2}{c}{Politicians} & \multicolumn{2}{c}{Audience} \\
 & Survey & Facebook & Twitter & Facebook & Twitter \\ \hline
\emph{Known topics from the survey} \\ \hline
Labor Market & 19.1 & 4.9 & 6.3 & 8.3 & 6.3 \\
General Social Policy & 12.9 & 1.1 & 1.5 & 1.4 & 1.4 \\
Currency \& Euro & 12.5 & 1.6 & 2.6 & 1.9 & 2.2 \\
Education & 7.2 & 1.1 & 1.4 & 0.5 & 1.1 \\
Economy & 7.0 & 0.5 & 0.6 & 0.5 & 0.6 \\
Infrastructure & 6.8 & 3.0 & 5.7 & 1.5 & 5.1 \\
Health Care \& Pensions & 5.7 & 0.9 & 1.0 & 0.5 & 0.7 \\
Migration \& Integration & 4.0 & 0.2 & 0.2 & 0.1 & 0.3 \\
Polity I & 4.0 & 0.4 & 0.3 & 0.3 & 0.2 \\
Family Policy & 3.3 & 2.4 & 2.4 & 2.3 & 3.0 \\
Law \& Order & 3.3 & 3.9 & 7.5 & 3.9 & 9.4 \\
Foreign Policy (Defense) & 2.9 & 2.4 & 3.4 & 2.3 & 3.2 \\
Budget \& Debt & 2.7 & 0.2 & 0.2 & 0.1 & 0.1 \\
Taxes & 2.6 & 0.2 & 0.3 & 0.1 & 0.2 \\
Foreign Policy (Europe) & 2.2 & 0.1 & 0.1 & 0.2 & 0.1 \\
General Fiscal Policy & 1.7 & 0.1 & 0.1 & 0.0 & 0.1 \\
Environment & 1.5 & 0.1 & 0.1 & 0.0 & 0.1 \\
Politics & 0.6 & 3.8 & 0.8 & 1.3 & 0.7 \\ \hline
\emph{New topics found on social media} \\ \hline
Campaigning (Local) &  & 21.2 & 13.7 & 5.6 & 7.4 \\
Campaigning (Events) &  & 21.1 & 12.4 & 1.9 & 4.6 \\
Political Debates &  & 8.5 & 13.6 & 21.1 & 18.8 \\
Polity II &  & 5.7 & 7.2 & 22.0 & 13.2 \\
Coalition Formation &  & 5.6 & 6.7 & 12.0 & 8.1 \\
Post Election &  & 3.8 & 3.8 & 6.9 & 4.8 \\
Parliamentary Procedures &  & 3.3 & 3.2 & 0.5 & 1.4 \\
Demonstrations &  & 2.3 & 2.8 & 0.5 & 2.0 \\
NSA Surveillance &  & 0.4 & 1.0 & 0.4 & 3.1 \\
Misconduct &  & 0.3 & 0.5 & 0.3 & 1.4 \\ \hline
\small{Note: Ranked by first given column.}
    \end{tabular}
\end{table}

Table \ref{tab:fraction} shows topic saliences in the five corpora (in percent). The first clear finding is that core policy areas like \emph{Labor Market} and \emph{General Social Policy} have a much higher salience in the survey than on social media. This resembles the results of \citet{Althaus.2002} who showed that readers of the paper version of the New York Times were more likely to expose themselves to ``public affairs coverage'' than the users of its online version.
Among policies, \emph{Infrastructure}, \emph{Family Policy} and \emph{Foreign Policy (Defense)} are all discussed frequently by politicians and audiences on social media. Aspects regarding NSA surveillance in the topic \emph{Law \& Order} are clearly overrepresented in messages by the audience and politicians on Twitter (see also \citealp{Jungherr.2016bb}). The latter finding demonstrates that the text analysis model performs well in allocating social media messages to known topics from the survey. Besides the discussed policy areas, politicians and audiences alike mostly address different topics than the ones salient in the survey, which lends support to H1.

It also becomes apparent that politicians use Facebook and Twitter in different ways. Their summed share of messages related to both \emph{Campaigning} topics is 42.3\% on Facebook compared to 26,1\% on Twitter.
In contrast, \emph{Political Debates} take up a higher share of politicians' tweets, and the latter medium is also used more extensively to discuss various policies like \emph{Infrastructure} and \emph{Law \& Order}. This indicates that politicians tailor their messages to different media logics and audiences. In line with our medium specific hypotheses, candidates use Twitter for the commentary of policies and unfolding public events (H3a), while trying to mobilize Facebook users to attend campaign events (H3b). 

Political audiences use social media overwhelmingly for \emph{Political Debates}, to scrutinize the relationship between state and citizens (\emph{Polity II}) and comment on \emph{Coalition Formation}. While politicians and audiences are more in sync on Twitter in that regard, there is a considerable disconnect between politicians and their audiences on Facebook. Although politicians devoted 42.3\% of their messages to campaigning, their audiences mostly talked about other topics.

\begin{table}
\centering
\caption{Rank correlations of topic salience in all corpora}
\label{tab:rankcorr}
\begin{tabular}{lcccc}
 & Survey & Facebook & Twitter & Facebook \\
 &  & politicians & politicians & audience \\ \hline
  \\[-0.2cm] 
  Facebook politicians &  0.43\\ 
  Twitter politicians &  0.59$^{* }$   &  0.95$^{***}$  &  \\ 
  Facebook audience &  0.53$^{* }$   &  0.91$^{***}$ &  0.90$^{***}$ &  \\ 
  Twitter audience &  0.58$^{* }$   &  0.89$^{***}$ &  0.95$^{***}$ &  0.92$^{***}$  \\ 
   \hline
   \\[-0.2cm] 
  \multicolumn{5}{p{10.5cm}}{\small{Note: Spearman's rho. For pairs including the survey, N=$18$. For social media pairs, N=$28$. $^{***}p<0.001$, $^{**}p<0.01$, $^*p<0.05$}}
\end{tabular}
\end{table}

A systematic way to analyze topic salience at the document level is to correlate the ranks of topics in each corpus using Spearman's rho. Several findings can be inferred from Table \ref{tab:rankcorr}. First, comparing the topic ranks in the survey with social media reveals varying results. Topic saliences in messages by politicians on Twitter ($p=0.011$), the Facebook audience ($p=0.024$) and the Twitter audience ($p=0.012$) are all rather similar to topic salience among a mass audience. Second, topic ranks of all social media corpora are nevertheless correlated more strongly with other social media corpora than with the survey ($p<0.001$).\footnote{The correlations between all social media corpora are very similar and highly significant ($p<0.001$) when only using the 18 known topics from the survey as input.} Third, correlations between the topic ranks in the survey and politicians' Facebook posts are weaker than in case of the other social media corpora ($p=0.075$). This lends support to H2 which postulates that candidates address topics relevant to a mass audience on Twitter while discussing such topics more sparsely on Facebook.  

We learn from these distributions that the messages of social media users are shaped by considerable mediation effects, which provides further evidence for H1. On social media, a specific subset of politically engaged citizens discusses specific topics via specific sociotechnical transmission mechanisms. Politicians seem to adjust to these mediated environments and thus have remarkably similar topic ranks like the personal networks they are most directly exposed to. They adopt the public communication practices of the Twittersphere \citep{Wu.2011} and use the medium primarily for political commentary, while trying to mobilize their interested followers on Facebook for campaign purposes. At the same time, correlations between the different content layers on social media -- except for the Facebook politicians corpus -- and the survey are still strong. Therefore, the stark differences in the sizes of topics per corpus revealed in Table \ref{tab:fraction} are reduced considerably when comparing the ranks of topics. When politicians and audiences talk about policies on social media, they tend to prioritize similar topics like respondents in a representative survey. This indicates that the public agenda is still rather cohesive during election campaigns -- independent of the medium.    

\subsection{Analysis at the word level}

The analyses at the document level revealed the salience of political topics in different media and content layers. But with the textual data at hand, we can move our analysis to the word level, which serves two purposes. First, it adds methodological robustness to the previous results. Given different types of media, it is to be expected that the language encountered in corpora is distinct, since specific conventions and space restrictions apply to interview situations, Facebook messages or tweets. The following analysis will rule out that differences in salience are mere artifacts of platform specific talk. Second, we can also investigate how different topics are perceived and talked about by a mass audience, politicians and their social media audiences. By that, we move beyond the salience of topics toward an identification of similar perspectives regarding political topics in different corpora. 

\begin{figure}[t]
	\centering	
	\includegraphics[width=1\textwidth]{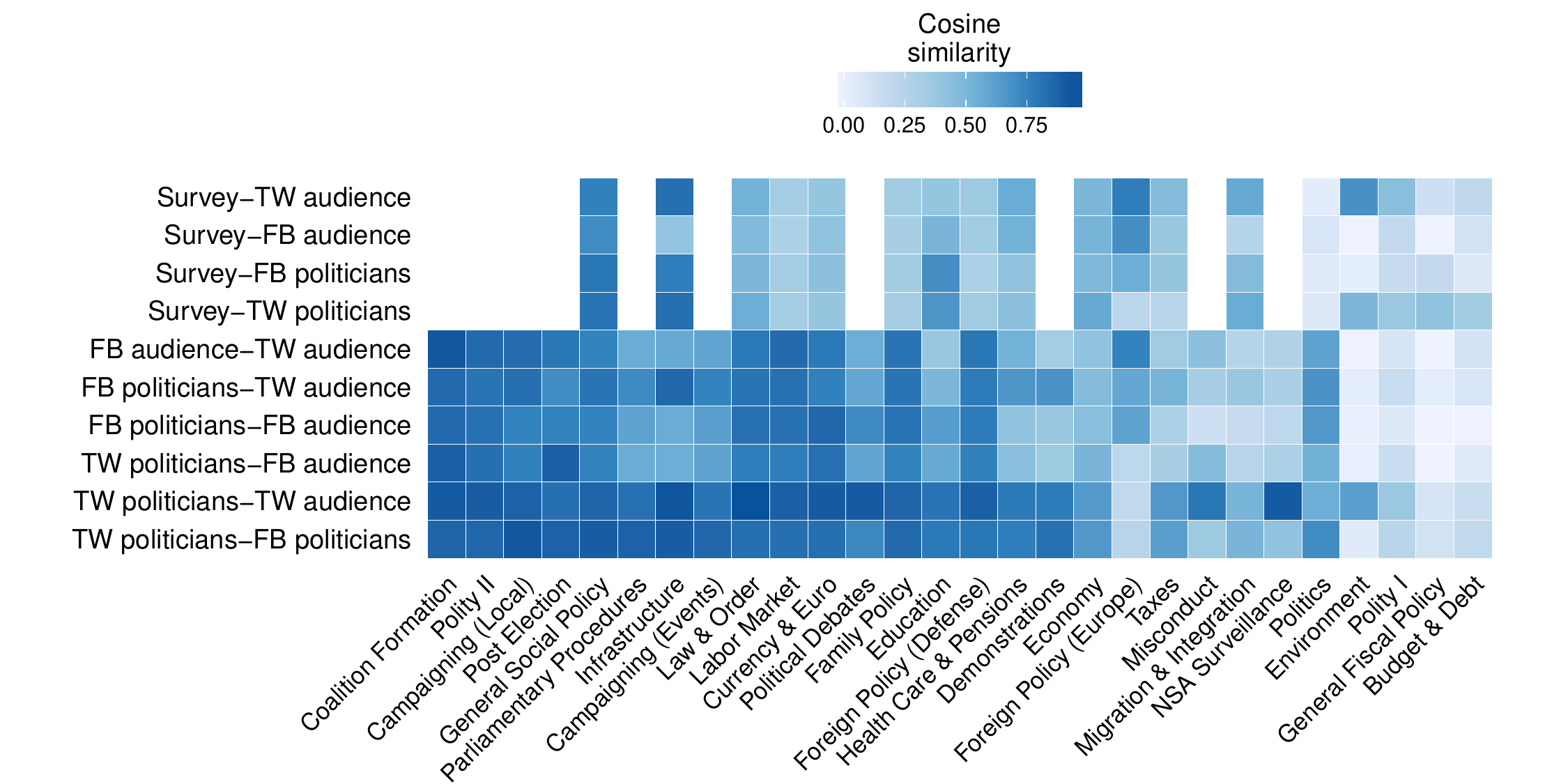}
	\caption{Cosine similarities between corpora by topic.}
	\label{fig:cosine}
\end{figure}

Since the underlying vocabulary in our model is the same across all five corpora, we can compare the topic-word distributions across media and content layers. For this, we calculate cosine similarities (scale 0 to 1) between all corpus pairs in each topic. Figure \ref{fig:cosine} displays the results, organized in decreasing order by the average cosine similarity per topic. The darker a cell, the more similarly a corpus pair discusses a topic. Cells of social media specific topics remain blank for the survey since the topics are not featured in this corpus. It becomes clear from the visualization that similarities between corpora vary considerably depending on the topic. 

In our final analysis, we take the values of the 240 cells in Figure \ref{fig:cosine} as the dependent variable in Ordinary Least Squares regressions.\footnote{Since most models were heteroscedastic, we use robust standard errors.} This allows us to assess if our hypotheses regarding similarities between different media and content layers hold at the level of words used in political communication. As independent variables, we construct a dummy indicating that the survey is part of a given corpus pair and, similarly, if the politicians' Facebook or Twitter corpus is featured in a corpus pair. We also take control variables into account that could influence the relationships between cosine similarities and the independent variables. First, we include a logged variable counting the combined number of tokens (the aggregate word count) in each corpus pair per topic. By that we take the size of corpus pairs into account which could systematically affect cosine similarities. We added dummies indicating that a corpus pair is from the same social medium, i.e., Facebook or Twitter, or that the group (politicians, audiences) in a corpus pair is the same. Moreover, we use the information from GLES on topic types, i.e., whether a topic belongs to the categories policy, politics or policy and classified the ten new social media topics analogously. Based on that, we construct a dummy signaling that a topic is related to politics aspects, which we assume are discussed more heterogeneously than the polity or policies.\footnote{This is not a hard distinction but rather an exploratory application of the GLES labels. In fact, it is difficult to distinguish policy aspects from politics aspects related to an issue.} Finally the 10 new topics identified on social media are marked by a dummy variable to account for differences in topic origin (\emph{New topics}).

\begin{table}
\centering
\caption{Models of cosine similarities between corpus pairs}
\label{tab:regression}
\begin{tabular}{l c c c c c }
	\hline
			& (1) & (2) & (3) & (4) & (5)\\
\hline
Survey       & $-0.12^{***}$ & $-0.12^{***}$ & $-0.10^{**}$  & $-0.11^{***}$ & $-0.09^{**}$  \\
             & $(0.03)$      & $(0.03)$      & $(0.03)$      & $(0.03)$      & $(0.03)$      \\
Facebook politicians      &               & $0.00$        & $0.02$        &               &               \\
             &               & $(0.02)$      & $(0.02)$      &               &               \\
Twitter politicians      &               &               & $0.05^{*}$    &               & $0.06^{**}$   \\
             &               &               & $(0.02)$      &               & $(0.02)$      \\
Twitter audience &               &               &               & $0.03$        & $0.05^{*}$    \\
             &               &               &               & $(0.02)$      & $(0.02)$      \\
Number of tokens (logged)  & $0.14^{***}$  & $0.14^{***}$  & $0.14^{***}$  & $0.14^{***}$  & $0.14^{***}$  \\
             & $(0.01)$      & $(0.01)$      & $(0.01)$      & $(0.01)$      & $(0.01)$      \\
Same medium & $0.08^{**}$   & $0.08^{**}$   & $0.08^{**}$   & $0.08^{**}$   & $0.08^{**}$   \\
             & $(0.03)$      & $(0.03)$      & $(0.03)$      & $(0.03)$      & $(0.03)$      \\
Same actor  & $0.05$        & $0.05$        & $0.05$        & $0.05$        & $0.05$        \\
             & $(0.03)$      & $(0.03)$      & $(0.03)$      & $(0.03)$      & $(0.03)$      \\
Topic type (politics=1)         & $-0.10^{*}$   & $-0.10^{*}$   & $-0.10^{*}$   & $-0.10^{*}$   & $-0.10^{*}$   \\
             & $(0.05)$      & $(0.05)$      & $(0.05)$      & $(0.05)$      & $(0.04)$      \\
New topics  & $-0.02$       & $-0.02$       & $-0.02$       & $-0.02$       & $-0.01$       \\
& $(0.05)$      & $(0.05)$      & $(0.04)$      & $(0.05)$      & $(0.04)$      \\
Constant  & $-0.59^{***}$ & $-0.59^{***}$ & $-0.61^{***}$ & $-0.60^{***}$ & $-0.62^{***}$ \\
             & $(0.07)$      & $(0.07)$      & $(0.07)$      & $(0.07)$      & $(0.07)$      \\
\hline
R$^2$        & 0.63          & 0.63          & 0.63          & 0.63          & 0.64          \\
Adj. R$^2$   & 0.62          & 0.61          & 0.62          & 0.62          & 0.63          \\
N    & 240           & 240           & 240           & 240           & 240           \\
\hline
\multicolumn{4}{p{8cm}}{\small{Note: OLS with robust standard errors in parentheses. $^{***}p<0.001$, $^{**}p<0.01$, $^*p<0.05$}}
\end{tabular}
\end{table}

Table \ref{tab:regression}, Model 1 shows that the corpus pairs including the survey have an approximately 12\% lower cosine similarity, other things being equal. This demonstrates that in addition to the differences in topic salience identified before (H1), the description of political problems in the survey is also markedly different from the average topically related social media message. In Model 2 and Model 3, we see that contrary to their posts on Facebook, the language in tweets by politicians is significantly more similar to other corpora than the average corpus pair. This means that on Twitter, candidates emphasize aspects of topics important to survey respondents, but also use language similar to other content layers on social media. Such a hybrid communication strategy by politicians can be interpreted as a synergy of mass communication strategies and messages targeted at individual follower networks on social media, much in line with the notion ``masspersonal'' ascribed to Twitter \citep{Wu.2011}.

Several more factors could influence these results. First, throughout Table \ref{tab:regression} the number of underlying tokens is highly correlated with the cosine similarity between the corpus pairs. This is to be expected by the law of large numbers as the empirical word distributions in both corpora become more stable with an increasing number of tokens and cosine similarities increase. Controlling for the size effect ex post in a multivariate regression model allows for a systematic comparison of similarities between corpus pairs. Second, politicians' tweets might reflect a typical communication style on Twitter and thus not be indicative of strategic considerations. To test this, we include a Twitter audience dummy in Model 4 without finding a significant effect. When including the Twitter audience dummy in combination with the Twitter politicians dummy in Model 5, it becomes significant, but the Twitter politicians variable still has superior explanatory power.\footnote{We do not include all corpus dummies at once because as part of corpus pairs, the dummies are not mutually exclusive. Therefore, we cannot exclude one dummy as the reference category and interpret results accordingly, as it is usually done with multiple-category variables.} Third, there is nonetheless evidence for platform effects, as \emph{Same medium} is a significantly positive predictor ($p\le0.009$ in all models). This demonstrates that politicians and audiences not only emphasize similar topics (H1) but also address similar aspects when talking about a topic on the same social media platform. Fourth, the results also hold when controlling for the insignificant variable \emph{Same actor}, which means that audiences and politicians adopt distinct communicative practices when using different media. Fifth, the topics primarily related to politics are more semantically diverse than the topic types polity and policy. Finally, similarities do not differ systematically depending on whether a topic is known from the survey or newly created from social media (\emph{New topics}). This is of particular methodological importance, as it demonstrates that semantic differences and the measurement of topic salience at the document level are not just artifacts of stark differences in the two communication situations interview and social media message. Finding a systematic effect of the two data generation modes would have questioned the validity of applying a model based on survey responses to social media data.\footnote{The main results are robust when rerunning the regression models using only known topics from GLES (with N=180 cells in Figure \ref{fig:cosine}) and also when solely taking the new social media topics into account (N=60 cells). The exception is the model with only known topics, in which the Twitter politicians and audience dummies lose their significance. Since some of the dummy variables overlap considerably, we also ran broader and smaller models with varying constellations of included variables, which confirm the main findings.} 

In sum, Table \ref{tab:regression} reveals similarities in political communication between corpora that resemble the ranks of topic salience at the document level. Still, problem descriptions in the survey and social media are more distinct than to be expected from Table \ref{tab:rankcorr}. This shows that overlaps in topic salience can still mask \emph{how} topics are talked about and perceived by different audiences.\footnote{Although agenda-setting is outside the scope of our study, this distinction resembles the conceptualization of first- and second-level effects found in the related literature \citep{McCombs.1997}.} We also learn that the distinct uses of Facebook and Twitter by politicians are not just byproducts of media specific language, but are rather driven by the strategic considerations we discuss in our theory. Moreover, the analysis at the word level makes the topic generation process more transparent and increases our understanding of mediation processes in political communication.

\section{Conclusion}
This study provides insights into how political communication is shaped by social media. We compared the topics of most importance to a mass audience in a representative survey to the topics discussed on Facebook and Twitter. Based on a text analysis model developed specifically for this purpose, we show that politicians and their audiences discuss different topics on social media than those salient among a mass audience. Moreover, politicians use Facebook and Twitter for different purposes which we relate to the distinct target groups candidates encounter. Taken together, our findings suggest that campaign strategies and political communication in general are mediated by varying sociotechnical affordances of social media platforms. 

These results challenge previous findings in political communication which indicated that politicians use the web rather conservatively and in a non-interactive manner \citep{Druckman.2010,Stromer.2000,Lilleker.2011,Gibson.2014,Larsson.2015}. However, our study does not necessarily contradict previous research, as each Internet application has specific affordances and politicians demonstrate a strategic awareness of various communication arenas. They complement the ``masspersonal'' communication \citep{Wu.2011} in the quasi-public sphere of Twitter with the more direct communication practices on Facebook for organizational and mobilization purposes. Our cross-media study also shows that relevant differences in political communication exist between social media platforms. This underscores the need to argue with the utmost caution when trying to infer findings from one platform to ``social media'' as a whole, as it has often been done.

Several findings indicate that the high-choice media environments of social media contribute to a fragmentation of the mass audience (\citealp{Althaus.2002}; \citealp{Prior.2007}; \citealp[p.~140]{Mcquail.2010}). Politicians and their audiences talk about policies sparsely, but rather discuss campaign-related events and topics specific to social media. Their salient topics resemble each other remarkably, compared to the priorities of survey respondents. Furthermore, the language used indicates that other aspects of similar topics are addressed on social media than in survey responses. Those differences notwithstanding, when candidates and audiences actually discuss policies on Facebook and Twitter, they prioritize those similarly like respondents in a representative survey. In this regard, the public agenda is still rather integrated. This points towards persistent -- although probably diffuse and mediated -- agenda-setting effects between mass media and social media \citep{Neuman.2014} as well as within social media. A fruitful research avenue would be to focus on the interactions of politicians and audiences with accounts of journalists and legacy media which bridge the gap between the general public and more particular sets of audiences on social media.

We also want to discuss the limitations of this study. The article focused on an election campaign with several high-attention events that were most commented-on on social media. A follow-up study during non-election periods could possibly reveal more issue-related communication in line with the preferences of a mass audience. Demographic information on the follower graphs of politicians on Facebook and Twitter would certainly enhance our findings, however, such data is not readily available to researchers. While our single-membership model is well suited for the short nature and narrow focus of social media messages, its downside is that the more multidimensional and textually rich messages necessarily also have to be allocated to just one topic. Moreover, due to restrictions in survey data containing an MIP and the considerable efforts required to mine social media data of all election candidates, we only investigated one campaign in one country. On the one hand, the mediation effects we found should travel well, as German social media campaigning was still in its infancy in 2013 and severe regulatory restrictions on micro-targeting apply \citep{Stier.2015c}. More sophisticated campaigns elsewhere might be better at addressing the particular needs of digital media. On the other hand, the rather ad hoc campaign style and personal use of accounts by several candidates in our case might be more suited for the peculiarities of social media than, for instance, the highly professionalized U.S. campaigns. Therefore, it remains to be seen to which extent the findings are applicable to online campaigning in other countries and the (near) future. 

The paper makes a methodological contribution to the field of political text analysis \citep{Grimmer.2013} by proposing a novel methodology to apply labeled text to a test corpus while also allowing for the introduction of additional topic categories. The method is applicable in a vast range of research contexts in political communication in which a test dataset deviates from a predefined coding scheme. Substantively, the paper shows that mediation effects induced by social media platforms and their sociotechnical environments are strongly felt in political communication. This means that social media is not an ideal data source for citizens seeking clearly structured information on policies or researchers using textual information to locate parties in an ideological space.

\section{Acknowledgements}
We thank three reviewers, the editors and participants at the ECPR General Conference 2016 for helpful comments.

\theendnotes

\bibliographystyle{apacite}
\bibliography{gles_bib}

\setcounter{table}{0}
\renewcommand{\thetable}{A\arabic{table}}

\setcounter{figure}{0}
\renewcommand{\thefigure}{A\arabic{figure}}

\begin{appendices}

\section{Appendix A: Description of the model}
\label{appendix:model}
\begin{figure}
	\centering
	\begin{center}
			\resizebox{12cm}{!}{
\begin{tikzpicture}


  \node[latent]			                                (theta) {$\boldsymbol{\theta}$};
  \node[const, left=of theta]	                        (alpha) {$\alpha$};

  \node[latent, below=of theta, yshift=+0.5cm]	        (z2) {$z^\mathcal{M}_{d}$};
  \node[obs, below=of z2]			                    (w2) {$\mathbf{w}^\mathcal{M}_{d}$};
  
  \node[obs,  right=of w2, xshift=-6.0cm]			    (w1) {$\mathbf{w}^\mathcal{G}_{d}$};
  \node[obs, above=of w1]		                        (z) {$z^\mathcal{G}_{d}$};

   \node[latent, left=of w2, xshift=-0.30cm]		    (phi) {$\boldsymbol{\phi}_k$};
   \node[const, above=of phi, yshift=-0.25cm]			(beta) {$\beta$};
   
  \coordinate[below=0.15cm of phi,yshift=+.05cm] (hidden);

  \edge {z} {w1};
  \edge {z2} {w2};
  \edge {phi} {w1};
  \edge {phi} {w2};
  \edge {alpha} {theta}; 
  \edge {beta} {phi}; 
  \edge {theta} {z};
  \edge {theta} {z2};

  \plate {data1} {(w1)(z)} {$\forall\; d \in [1,D^\mathcal{G}]$} ;
  \plate {data2} {(w2)(z2)} {$\forall\; d \in [1,D^\mathcal{M}]$} ;
  \plate {topics} {(phi)(hidden)} {$\forall\; k \in [1,\infty[$} ;

  \node[text width=2cm, below=of phi, xshift=-.4cm, align=center] 
    {topic-word distributions};
        
  \node[text width=3cm, right=of w2, xshift=-.65cm, align=left] 
    {words in social\\ media messages};
    
  \node[text width=3cm, right=of z2, xshift=-.6cm, align=left] 
    {unknown\\ topic assignment};
    
  \node[text width=3cm, right=of theta, xshift=-.6cm, align=left] 
    {global topic\\ popularity};

  \node[text width=3cm, left=of z, xshift=0.3cm, align=left] 
    {observed\\ topic assignment};
    
  \node[text width=2.5cm, left=of w1, xshift=-0.2cm, align=left] 
    {words in GLES responses};

\end{tikzpicture}
}
	\end{center}	
  	\caption{\textbf{Model in Plate Notation:} Random variables are represented by nodes. Blank nodes are used for the (unknown) hidden variables $\boldsymbol{\theta}$, $\boldsymbol{\phi}$ and $\mathbf{z}^\mathcal{M}$. Shaded nodes denote the observed words $\mathbf{w}^\mathcal{M}$ and $\mathbf{w}^\mathcal{G}$ as well as the observed labels $\mathbf{z}^\mathcal{G}$, bare symbols indicate the fixed priors $\alpha$ and $\beta$. Directed edges between the nodes then define conditional probabilities, where the child node is conditioned on its parents. The plate surrounding $\boldsymbol{\phi}_k$ indicates the replication of the node for each mixture component and the plates surrounding $\mathbf{w}^\mathcal{G}_{d}$ and $\mathbf{w}^\mathcal{M}_{d}$ indicate the replication of the nodes over the $\mathcal{D}^\mathcal{G}$ labeled and $\mathcal{D}^\mathcal{M}$ unlabeled documents, respectively.}
  \centering
  \label{fig:plate}
\end{figure}
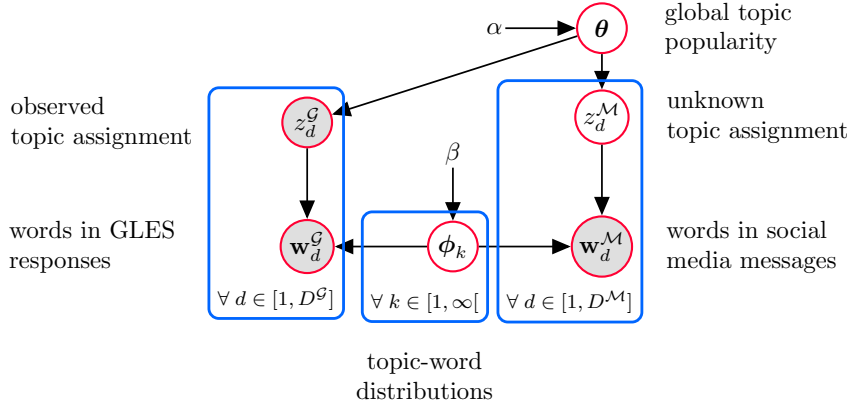

Following the common notation for probabilistic language models, we treat a GLES response $d$ as a tuple of a topic indicator variable $z^\mathcal{G}_d$ and a vector $\mathbf{w}^\mathcal{G}_{d}$ of observed words. The indicator $z^\mathcal{G}_d\in [1, ...,\hat{K}]$ is an assignment to one of $\hat{K}$ initial GLES topics, and each of the $n^\mathcal{G}_d$ observed words $w^\mathcal{G}_{di}$ in $\mathbf{w}^\mathcal{G}_{d}$ is chosen from a vocabulary of $V$ terms, leading to the collection of GLES responses $\mathcal{D}^\mathcal{G} = \{(\mathbf{w}^\mathcal{G}_d,z^\mathcal{G}_d), ..., (\mathbf{w}^\mathcal{G}_{D^\mathcal{G}},z^\mathcal{G}_{D^\mathcal{G}})\}$. 
Analogously, we treat a social media message $(\mathbf{w}^\mathcal{M}_{d},z^\mathcal{M}_d)$ as a tuple of words and a topic assignment. While the $n^\mathcal{M}_d$ words $\mathbf{w}^\mathcal{M}_{d}$ are chosen from the same vocabulary as in the case of GLES responses, the topic assignments $\mathbf{z}^\mathcal{M}$ are unobserved and not restricted to the initial $\hat{K}$ topics. The collection of social media messages is $\mathcal{D}^\mathcal{M} = \{(\mathbf{w}^\mathcal{M}_d,z^\mathcal{M}_d), ..., (\mathbf{w}^\mathcal{M}_{D^\mathcal{M}},z^\mathcal{M}_{D^\mathcal{M}})\}$. We refer to the collection of all documents (GLES responses and social media messages) by $\mathcal{D}= \mathcal{D}^\mathcal{G} \cup \mathcal{D}^\mathcal{M}$. The generative storyline for the documents in our model, i.e. the mechanism assumed to create responses and messages, can be described by the following steps.
\begin{enumerate}

\item For the unrestricted global topic popularity a distribution $\boldsymbol{\theta}$ is drawn from an infinite dimensional Dirichlet distribution (i.e., $K\to \infty$) with parameter $\alpha$:
\begin{align*}
    \boldsymbol{\theta} \sim Dir(\tfrac{\alpha}{K}) \mbox{ .}
\end{align*}
\item For each topic $k = [1,...,\infty[$, a distribution $\boldsymbol{\phi}_{k}$ over the vocabulary is drawn from a V-dimensional Dirichlet distribution parametrized by $\beta$:
\begin{align*}
    \boldsymbol{\phi_{k}} \sim Dir(\beta) \mbox{ .}
\end{align*}

\item The GLES responses $\mathcal{D}^\mathcal{G}$ and social media messages $ \mathcal{D}^\mathcal{M}$ are then generated by the same mechanism of first drawing a topic index $z_d$ from the topic popularity $\boldsymbol{\theta}$ and subsequently drawing multinomially distributed words $\mathbf{w}_d$ from a topic indexed by $z_d$:
\begin{align*}
    \mathbf{w}_d \sim Mult(\boldsymbol{\phi}_{z_d}) \mbox{ ,} &&& z_d \sim Cat(\boldsymbol{\theta}) \mbox{,}
\end{align*}

where we omitted the indexes $\mathcal{G}$ and $\mathcal{M}$ for readability, leading to the joint probability for the model 
\begin{align*}
    p(\mathcal{D}, \mathbf{z},\boldsymbol{\theta}, \boldsymbol{\phi}) = 
    &Dir( \boldsymbol{\theta} \mid \tfrac{\alpha}{K}) \\
    &\prod_{k=1}^{\infty}\, Dir( \boldsymbol{\phi}_k \mid \beta) \\
    &\prod_{d=1}^{|\mathcal{D}^\mathcal{G}|}\, 
    Mult(\mathbf{w}^\mathcal{G}_{d}\mid \boldsymbol{\phi}_{z^\mathcal{G}_{d}}) \,
    Cat(z^\mathcal{G}_{d} \mid \boldsymbol{\theta}) \\
    &\prod_{d=1}^{|\mathcal{D}^\mathcal{M}|}\, 
    Mult(\mathbf{w}^\mathcal{M}_{d}\mid \boldsymbol{\phi}_{z^\mathcal{M}_{d}}) \,
    Cat(z^\mathcal{M}_{d} \mid \boldsymbol{\theta})\mbox{ .}
\end{align*}
\end{enumerate}
Figure~\ref{fig:plate} gives a graphical representation of our model.

\section{Appendix B: Description of the inference}
\label{appendix:model_inference}
For inference, we resort to collapsed Gibbs sampling of the latent indicator variables $\mathbf{z}^\mathcal{M}$, with all other variables integrated out. The full conditional posterior probability for a topic $k$ is then given by
\begin{align*}
	p(z^\mathcal{M}_d = k) &\propto 
	\begin{cases} 
		  \dfrac{{}^{\neg d}n_k}{{}^{\neg d}n_. + \alpha} \;\; p(\mathbf{w}^\mathcal{M}_{d} \mid {}^{\neg d}\mathbf{w}^\mathcal{M}_{z=k}, \mathbf{w}^\mathcal{G}_{z=k}, \beta) & \text{if }  k \leq K \\[1em]
		  \dfrac{\alpha}{{}^{\neg d}n_. + \alpha} \;\; p(\mathbf{w}^\mathcal{M}_{d} \mid \beta) & \text{if } k = K + 1 \mbox{ ,}
	\end{cases}
\end{align*}
where $n_k=\sum_{d=1}^{D} \mathbbm{1}[z_d = k]$ is the number of documents and $\mathbf{w}_{z=k}$ are the words assigned to the $k^{th}$ topic. The superscript $\neg d$ indicates that the current document $d$ is excluded from consideration. We further use a approximation for the likelihood of message $d$ under topic $k$
\begin{align*}
	p(\mathbf{w}^\mathcal{M}_{d} \mid {}^{\neg d}\mathbf{w}^\mathcal{M}_{z=k}, \mathbf{w}^\mathcal{G}_{z=k}, \beta) 
	&= \int_{\phi_k} Mult(\mathbf{w}^\mathcal{M}_{d} \mid \boldsymbol{\phi}_k) \, Dir(\boldsymbol{\phi}_k \mid \hat{\beta}_k^{\neg d}) \; d\ \boldsymbol{\phi}_k \notag \\
	&\approx \prod_{i=1}^{n_d^\mathcal{M}}\dfrac{{}^{\neg d}n_{kw_{di}}^\mathcal{M} + n_{kw_{di}}^\mathcal{G} + \beta}{{}^{\neg d}n_{k.}^\mathcal{M} + n_{k.}^\mathcal{G} + V\beta} \mbox{ ,}
\end{align*}
with $n_{kw}$ being the number of times term $w$ is used in topic $k$. Likewise, the likelihood of $\mathbf{w}^\mathcal{M}_{d}$ being generated from a yet unseen topic is
\begin{align*}
	p(\mathbf{w}^\mathcal{M}_{d} \mid \beta) 
	&= \int_{\phi} Mult(\mathbf{w}^\mathcal{M}_{d} \mid \boldsymbol{\phi}) \, Dir(\boldsymbol{\phi} \mid \beta) \; d\phi \notag \\
	&\approx \dfrac{1}{V^{n_d^\mathcal{M}}} \mbox{ .}
\end{align*}
Given the priors $\alpha$ and $\beta$, Gibbs sampling in this model reduces to iteratively sampling the indicator variables $\mathbf{z}^\mathcal{M}$ from their full conditional posterior probability. Once a sufficient number of iterations is completed, the distribution of the indicator variables and as such the number of topics $K$ with associated observations is independent from its initialization and the sampler has converged. For a more in-depth discussion of Gibbs sampling for Dirichlet process mixture models the interested reader is referred to \cite{neal2000markov}.

\section{Appendix C: Survey dataset}
\label{appendix:survey}
Topics useful for the training of the model should be sufficiently discriminative and relatively equal in size. The Rolling-Cross-Section survey of the German Longitudinal Election Study (GLES) allows for answers to be coded as politics, polity, or one of 13 policy areas.
From the latter, three were removed (206 answers on ``other'' problems, 102 on East Germany, and 1 on cultural leisure policy).
The sizes of the resulting areas ranged from 188 (``politics'') to 6,010 (``social policy'') observations. For our purposes, some areas had to be split by testing different subtopic combinations and qualitatively assessing topic overlaps, while others could be used as they are.

In essence, six areas were kept as they are. ``Social policy'' was split into \emph{General Social Policy}, \emph{Family Policy}, \emph{Health Care \& Pensions} and \emph{Migration \& Integration}. ``Fiscal policy'' was split into \emph{General Fiscal Policy}, \emph{Currency \& Euro}, \emph{Taxes} and \emph{Budget \& Debt}. GLES topics on foreign policy and defense policy were merged into a training topic on \emph{Foreign Policy (Defense)}, but the general subtopic on Europe was extracted to stand alone. As a result, we obtained 18 discriminative topics (totalling 23,295 responses in one politics class, one polity class and 16 policy classes). The coding is reproducible from Table \ref{tab:topics_gles}.

\begin{table}[H]
    \centering
    \begin{threeparttable}
		\caption{Construction of training topics.}
		\label{tab:topics_gles}
		\begin{tabular}{llll}
Label & Observations & GLES Codes \\ \hline
Politics & 188 & 1XXX \\
Polity I & 1,063 & 2XXX \\
Budget \& Debt & 629 & 431X \\
Currency \& Euro & 2,830 & 433X \\
Economy & 1,618 & 39XX, 40XX \\
Education & 1,624 & 41XX \\
Environment & 352 & 36XX \\
Family Policy & 776 & 371X \\
Foreign Policy (Defense) & 701 & 310X, 312X, 316X, 317X, 318X, 329X, 330X, 331X, 332X, 333X, 339X \\
Foreign Policy (Europe) & 561 & 311X \\
General Fiscal Policy & 389 & 430X, 439X \\
General Social Policy & 2,950 & 370X, 372X, 373X, 378X, 379X \\
Health Care \& Pensions & 1,333 & 374X, 376X, 377X \\
Infrastructure & 1,580 & 35XX \\
Labor Market & 4,358 & 38XX \\
Law \& Order & 793 & 34XX \\
Migration \& Integration & 951 & 375X \\
Taxes & 599 & 432X \\ \hline
		\end{tabular}
		\begin{tablenotes}
      \small
      \item \small Note: Topic codes from \citet{GLES.2013.agendafragen}.
    \end{tablenotes}
  \end{threeparttable}
\end{table}

\end{appendices}
 
\end{document}